\documentclass[prb,twocolumn,showpacs]{revtex4}
\usepackage{graphicx}
\usepackage{amssymb}

\newcommand{\rr}{{\mathbf r}}
\newcommand{\RR}{{\mathbf R}}

\begin{document} 

\title{Rolling and sliding of a nanorod between two planes:
  Tribological regimes and control of friction}

\author{Mykhaylo Evstigneev} \email[E-mail:]
{mykhaylo@physik.uni-bielefeld.de}
\author{Peter Reimann}
\affiliation{Universit\"at Bielefeld, Fakult\"at f\"ur Physik, 33615
Bielefeld, Germany}

\begin{abstract}
The motion of a cylindrical crystalline nanoparticle sandwiched between two
crystalline planes, one stationary and the other pulled at a constant
velocity and pressed down by a normal load, is considered
theoretically using a planar model. The results of our model
calculations show that, depending on load and velocity, the
nanoparticle can be either rolling or sliding. At sufficiently high
normal loads, several
sliding states characterized by different friction forces can coexist,
corresponding to different orientations of the nanoparticle, and
allowing one to have low or high friction at the same
pulling velocity and normal load.
\end{abstract}



\maketitle
\section{Introduction}
\label{secI}
Because of its great importance for nanotechnological applications,
control of friction at the nanoscale is a hot topic of current
research.  While conventional lubricants cannot be applied for this
purpose, research efforts have been guided by the vision of Richard Feynman
that the nanobearings can ``run dry'' \cite{Feynman}. It has been
shown that decreasing the normal load \cite{Socoliuc04} or pulling
velocity \cite{Krylov05}, as well as normal load actuation
\cite{Socoliuc07} can lead to a dramatic friction reduction. A
mechanism of friction control particularly pertinent to our present
work is the so-called structural lubricity, or superlubricity
\cite{Drexler87, Merkle93, Hirano, Verhoeven04, Muser03, Muser04,
Martin93, Ko00, Dienwiebel04}, arising due to the structural
incommensurability of the two contacting surfaces. More precisely,
each atom of the sliding surface feels the force generated by the
periodically arranged atoms of the substrate, so that the total
friction force is the sum of the forces felt by each atom of the
slider. If the slider and the substrate are incommensurate, these
forces add up randomly, resulting in nearly frictionless sliding. The
phenomenon of superlubricity has been observed experimentally
\cite{Hirano, Martin93, Ko00, Dienwiebel04}. On the other hand, it has
been demonstrated by means of molecular dynamics simulations
\cite{Depondt98}, as well as stochastic modelling and experiment
\cite{Filippov08} that a flat nanoobject (e.g. a graphite flake) in
contact with the surface quickly reorients itself into the
``commensurate'' state of high friction, even though the initial
orientation may be the superlubric one. 

Finally, one can influence the friction forces by using the rolling
motion of round nanoparticles in between the surfaces \cite{Braun,
Legoas04, Kang04, Ritter02, Falvo99, Falvo00}. Nanoparticle rolling can be
identified experimentally by making a small indentation on its surface
with a sharp tip of an atomic force microscope and subsequent
localization of this mark after the manipulation
\cite{Ritter02}. Alternatively, for a highly symmetric nanoparticle,
such as a carbon nanotube, a rolling regime can be identified from the
characteristic periodicity of the time-dependent friction force -- in
the rolling state, this periodicity is proportional to the
circumference of the nanoparticle \cite{Falvo99, Falvo00}.

\begin{figure}[h] 
\includegraphics[scale=0.29]{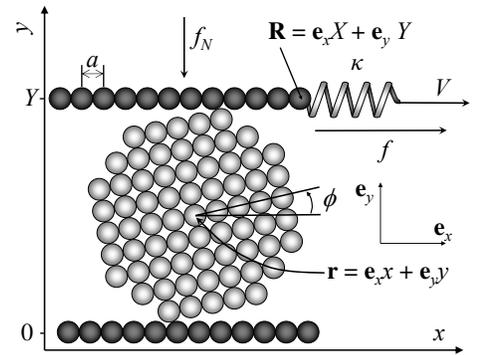}
\caption{Schematic illustration of the system:
  a nanoparticle is sandwiched between the stationary lower plane and
  the upper plane pressed down by the normal load $f_N$ and attached
  to a spring of stiffness $\kappa$, whose other end is pulled with
  the velocity $V$. The friction force $f$ is deduced from the elastic
  deformation of the spring attached to the upper plane,
  cf. Eq.~(\ref{300}).}
\label{fig1}
\end{figure}

In this paper, we consider the motion of a cylindrical nanoparticle
sandwiched between two planes, see Fig.~\ref{fig1}. Because of the
unavoidable deviations from a perfect rotational symmetry, the
surface of the nanoparticle actually consists of facets characterized
by different lattice constants, and hence by different
commensurabilities with both planes \cite{Falvo00}.  Based on a simple model
described in Section~\ref{secII}, we show in Section~\ref{secIII} that
the nanoparticle can be stabilized in several friction states, namely,
rolling friction and different sliding regimes that can be realized
depending on which facets of the nanoparticle are in contact with the
upper and lower surface. These results are summarized in a state
diagram showing the stability regions of different friction states
depending on the two experimental control parameters -- normal load
and pulling velocity. The state diagram contains a region, where
sliding friction regimes characterized by commensurate contact (high
friction forces) and incommensurate contact (low friction forces) of
the nanoparticle coexist. We propose a friction switching scenario,
allowing one to realize either low or high friction forces at the same
pulling velocity and normal load.

\section{The Model}
\label{secII}
\subsection{Equations of motion}
We consider a crystalline nanoparticle, approximating as close as
possible a round shape of some preset radius, see
Fig.~\ref{fig1}. Unlike the symmetric fullerene-like molecules with
equivalent `facets' studied in Refs.~\onlinecite{Braun, Legoas04,
Kang04}, in our system, the facets have different number of atoms and
different commensurability with the planes, see Fig.~\ref{fig1}, so
that, depending on which facets are in contact with the planes,
different friction regimes may be realized.  While our model system
from Fig.~\ref{fig1} is two-dimensional, the main results immediately
carry over to some three-dimensional objects, such as nanorods. We
henceforth consider a single particle, expecting analogous findings
for the case of many particles, provided they are sufficiently dilute.

The overall geometric configuration of the system is completely
determined by the nanoparticle's center of mass $\rr = x\mathbf e_x +
y\mathbf e_y$, its rotation angle $\phi$, and the position $\RR =
X\mathbf e_x + Y\mathbf e_y$ of some reference atom from the top
plane, which we assume to be horizontal at all times; here $\mathbf
e_{x,y}$ are the unit vectors in the $x$- and $y$-direction. If the
nanoparticle and the planes could be viewed as rigid bodies, the
equations of motion for the generalized coordinates $(\rr, \phi,
\RR)$ would be derivable from the Lagrangian
\begin{eqnarray}
&&L(\rr, \phi, \RR, \dot\rr, \dot\phi, \dot\RR) =
\frac{m\dot\rr^2}{2} + \frac{I\dot\phi^2}{2} + \frac{M\dot\RR^2}{2}
- U_B(\rr,\phi) \nonumber\\
&&\ \ - U_T(\RR-\rr, \phi) - \frac{\kappa(X -
  Vt)^2}{2} + f_N Y\ ,
\label{10}
\end{eqnarray}
where the first two terms represent the translational and rotational
kinetic energy of the nanoparticle with mass $m$ and moment of inertia
$I$; the third term describes the kinetic energy of the upper plane of
mass $M$; the fourth and the fifth terms correspond to the interaction
between the nanoparticle and the bottom and top planes, respectively;
the last two terms correspond to the energy of elastic deformation of
the spring of stiffness $\kappa$ whose other end is pulled at the
constant velocity $V$, and the energy of the upper plane produced by
the normal load. Therefore, the elastic force generated by the spring
is equal in magnitude to the instantaneous friction force
\begin{equation}
f = -\kappa(X - Vt)\ .
\label{300}
\end{equation}
We furthermore assume that the separation between the two planes --
the nanoparticle's diameter -- is sufficiently large, so that the
interaction energy between them is negligible.

In reality, the atoms of the nanoparticle and the planes are not
rigidly coupled to each other, and their motion affects the motion of
the global coordinates $\rr$, $\phi$, and $\RR$. If the time scales
of the overall nanoparticle motion is much slower than the time scale
of individual atoms, the interaction between the global degrees of
freedom, $\rr, \phi, \RR$, and those of atoms composing the
nanoparticle, both planes, and the spring can be approximately taken
into account by means of the following three modifications of the
equations of motion generated by the Lagrangian (\ref{10})
\cite{Evst10}:

(i) renormalization of the forces and torques acting on the relevant
coordinates $\rr$, $\phi$, and $\RR$;

(ii) introduction of velocity-dependent dissipation forces
describing the effect of energy loss from the
global degrees of freedom $\rr, \phi$, and $\RR$ into the atomistic
degrees of freedom of the nanoparticle, both planes, and the spring
attached to the upper plane;

(iii) introduction of noise corresponding to the inverse process of energy
transfer from random atomic vibrations into the global degrees of
freedom.

The effect (i) is accounted for by reinterpreting the energies
$U_{B,T}$ as free energies. The dissipative forces (ii) due to the internal
degrees of freedom of the nanoparticle and the planes can be derived from the
dissipation function \cite{Landau},
\begin{equation}
\Phi(\rr, \phi, \RR, \dot\rr, \dot\phi, \dot\RR) =
\frac{1}{2}\sum_{q, q'}\eta_{qq'}(\rr, \phi, \RR)\,\dot q\dot q'\ ,
\label{20}
\end{equation}
where the generalized coordinate indices $q, q'$ run over the values
$(x,y,\phi, X, Y)$, and the dissipation coefficients are symmetric,
\begin{equation}
\eta_{qq'} = \eta_{q'q}\ .
\label{30}
\end{equation} 
These forces are written as velocity derivatives of the dissipation
function, $f_q^{diss} = -\partial\Phi/\partial\dot q =
-\sum_{q'}\eta_{qq'}\dot q'$. Similarly, an additional dissipative
force on the upper plane arises due to the internal degrees of freedom
of the spring: $f_{spring}^{diss} = -\eta_S(\dot\RR - V\mathbf
e_x)$. This additional dissipation channel is not present in the
dissipation function (\ref{20}), because the position $Vt$ of the pulled
end of the spring is not a generalized coordinate in the Lagrangian
(\ref{10}); therefore, the dissipative force $f_{spring}^{diss}$
should be included ``by hand''. Finally, we account for the
noise effect (iii) by adding suitably chosen Gaussian white noises to
the right-hand side of the equations of motion:
\begin{eqnarray}
&&m\ddot\rr = -\nabla [U_B(\rr, \phi) - U_T(\RR-\rr, \phi)]
  \nonumber\\ &&\ \ \ -\eta_{\rr\rr}\dot\rr -
  \eta_{\rr\phi}\dot\phi - \eta_{\rr\RR}\dot\RR +
  \xi_\rr(t),\nonumber \\ &&I\ddot\phi = -\frac{\partial[U_B(\rr,
  \phi) + U_T(\RR-\rr, \phi)]}{\partial\phi} \nonumber\\ &&\
  \ \ -\eta_{\phi\rr}\dot\rr - \eta_{\phi\phi}\dot\phi -
  \eta_{\phi\RR}\dot\RR + \xi_\phi(t),\nonumber \\ &&M\ddot\RR =
  -\nabla U_T(\RR-\rr, \phi) - \kappa(X-Vt)\mathbf e_x - f_N\mathbf
  e_y \nonumber\\ &&\ \ \ -\eta_{\RR\rr}\dot\rr -
  \eta_{\RR\phi}\dot\phi - \eta_{\RR\RR}\dot\RR - \eta_S(\dot\RR
  - V\mathbf e_x) \nonumber\\
&&\ \ \ \ \ \ \ \ \ \ + \xi_\RR(t) + \xi_S(t)\ .
\label{40}
\end{eqnarray}
Here, $\eta_{\rr\rr}$ is a tensor with components $\eta_{xx},
\eta_{xy}, \eta_{yx} = \eta_{xy}, \eta_{yy}$, with similar definitions
for $\eta_{\rr\RR}$ and $\eta_{\RR\RR}$. Also, $\eta_{\rr\phi}$ and
$\eta_{\RR\phi}$ are vectors, e.g. 
$\eta_{\rr\phi} = \eta_{x\phi}\mathbf e_x +
\eta_{y\phi}\mathbf e_y$ with a similar definition for
$\eta_{\RR\phi}$, and $\eta_{\phi\rr} = \eta_{\rr\phi}^T$,
$\eta_{\phi\RR} = \eta_{\RR\phi}^T$ are the respective
transposed vectors. The Gaussian white noises $\xi_q(t)$, $q = x, y,
\phi, X, Y$ have zero
mean and obey the fluctuation-dissipation theorem of the second kind,
$\langle \xi_q(t)\xi_{q'}(t')\rangle =
2T\eta_{qq'}(\rr,\phi,\RR)\delta(t'-t)$. The noise $\xi_S(t)$ due to the
spring is uncorrelated with the noises $\xi_q(t)$ and its
autocorrelation function is $\langle\xi_{S\alpha}(t)\xi_{S\beta}(t')\rangle =
2T\eta_S\delta_{\alpha\beta}\delta(t'-t)$, where $\alpha, \beta$ refer
to the coordinates $x, y$. 

\subsection{Choice of the functional form for the potentials and
  dissipation coefficients}
Having written down the general equations of motion (\ref{40}) for the
relevant coordinates, we should specify the functional forms of the
potentials $U_{B,T}$, as well as the dissipation coefficients
$\eta_{\rr\rr}$, $\eta_{\rr\phi}$, etc. We consider the potentials
first. For simplicity, we assume the upper and the lower planes to be
equivalent, leading to
\begin{equation}
U_T(\RR-\rr, \phi + \pi) = U_B(\rr, \phi) =: U(\rr, \phi)\ ,
\label{50}
\end{equation}
and leaving us with the necessity to determine the function $U(\rr,
\phi)$. This function should possess the
symmetries
\begin{equation}
U(\rr,\phi) = U(\rr + a\mathbf e_x, \phi) = U(\rr, \phi +
2\pi/N)\ , 
\label{60}
\end{equation}
where $a$ is the lattice constant of the plane, and $N$ is an integer
related to the rotational symmetry of the particle. For example, for a
nanoparticle as sketched in Fig.~\ref{fig1}, we have $N = 4$. There
are many functions with the property (\ref{60}). To make a physically
motivated choice, we view the nanoparticle as a collection of
periodically arranged ``pseudoatoms'', where each pseudoatom
represents a group of closely arranged real atoms of the particle. In
particular, if Fig.~\ref{fig1} depicts a cross-section of a nanorod,
then each pseudoatom describes the cumulative effect of an atomic row
along the axis perpendicular to the plane of the figure. The $i$th
pseudoatom of the nanoparticle has the coordinate
\begin{equation}
\mathbf{r}_i(\rr, \phi) = \mathbf{r} + d_i\mathbf e_i(\phi)\ ,
\label{70}
\end{equation}
where $d_i = |\mathbf{r}_i - \mathbf{r}|$ is the rigidly fixed
distance of the $i$th pseudoatom from the center of mass, and $
\mathbf e_i(\phi) = \mathbf e_x \cos(\theta_i + \phi) + \mathbf
e_y \sin(\theta_i + \phi)$
is the unit vector pointing from the center of mass to the $i$th pseudoatom.
The constant angles $\theta_i$ are completely determined by the crystal
structure of the nanoparticle. 

The potential of interaction with the bottom plane is the sum of the
respective interaction energies of all pseudoatoms
\begin{equation}
U(\rr,\phi) = \sum_i u(\rr_i(\rr,\phi))\ .
\label{80}
\end{equation}
The functions $u(\rr) = u(\rr + a\mathbf e_x)$ can be expanded into
Fourier series in $x$ with the expansion coefficients depending on the
$y$-component of $\rr$. Neglecting the second and higher harmonics,
we employ the functional form
\begin{equation}
u(\rr) = u_0(y) + u_1(y)\cos\frac{2\pi x}{a}\ . 
\label{90}
\end{equation}
To account for the possibility of adhesion, the zero-order term is
taken to be the Lennard-Jones potential,
\begin{equation}
u_0(y) = \varepsilon
\left((\sigma/y)^{12} - 2(\sigma/y)^6\right)\ ,
\label{100}
\end{equation}
where $\varepsilon$ is the adhesion energy and $\sigma$ the equilibrium
separation from the surface.  

The function $u_1(y)$ in Eq.~(\ref{90}) has the physical
meaning of the corrugation amplitude of the potential (\ref{90}) in the
$x$-direction. We assume it to increase exponentially upon approaching
the surface:
\begin{equation}
u_1(y) = \Delta U e^{-(y-\sigma)/\lambda}\ ,
\label{110}
\end{equation} 
where $\Delta U$ is the corrugation at the equilibrium separation
$\sigma$, and $\lambda$ is the characteristic decay length.

With respect to the dissipation coefficients of the nanoparticle, we
should distinguish between the contributions due to the bottom and the
top planes. We assume that in the course of its translational and
rotational motion, the total dissipative force on the nanoparticle is
a sum of the respective contributions from all its pseudoatoms. That
is, the motion of the $i$th pseudoatom with the velocity $\dot\rr_i$
relative to the bottom plane
results in the dissipative force $-\eta(\rr_i)\dot\rr_i$ on that
pseudoatom. The expression for the dissipative force contribution from
the top plane is similar, but with the dissipation coefficient
$\eta(\RR-\rr_i)$ and the relative velocity $\dot \rr_i -
\dot\RR$. 

For simplicity, we assume isotropy of the damping
coefficients $\eta(\rr_i)$, which are treated as scalar functions of
the position $\rr_i$. We choose the following functional form:
\begin{equation}
\eta(\rr_i) = \eta_0 e^{-(y_i-\sigma)/\xi}\ ,
\label{120}
\end{equation}
$\eta_0$ being the damping coefficient at the minimum of the
Lennard-Jones potential (\ref{100}), and $\xi$ the decay length.
Explicitly, the velocity of $i$th pseudoatom is expressed in terms of
the generalized velocities $\dot\rr$ and $\dot\phi$ as
\begin{equation}
\dot\rr_i(\rr,\phi) = \dot\rr
+ d_i\mathbf t_i(\phi)\dot\phi\ ,
\label{130} 
\end{equation}
where the tangential vector is
\begin{equation}
\mathbf t_i(\phi) = \frac{d\mathbf e_i(\phi)}{d\phi} =
-\mathbf e_x\sin(\theta_i + \phi) + \mathbf
e_y\cos(\theta_i+\phi)\ . 
\label{140}
\end{equation}
The dissipative force due to the bottom plane is
\begin{equation}
\mathbf f^B_{diss} = -\sum_i\eta(\rr_i)\dot\rr_i =
-\sum_i\eta(\rr_i)\dot\rr - \sum_i\eta(\rr_i)d_i\mathbf
t_i\dot\phi\ .
\label{150}
\end{equation}
The dissipative force due to the top plane has a similar form, but
with the velocity $\dot\rr_i$ replaced with $\dot\rr_i - \dot\RR$ and
$\eta(\rr_i)$ with $\eta(\RR - \rr_i)$, that is
\begin{equation}
\mathbf f^T_{diss} =
-\sum_i\eta(\RR-\rr_i)(\dot\rr - \dot\RR) - \sum_i\eta(\RR-\rr_i)d_i\mathbf
t_i\dot\phi\ .
\label{160}
\end{equation}
Summing all the contributions, we obtain:
\begin{eqnarray}
&&\eta_{\rr\rr} = \sum_i[\eta(\rr_i) + \eta(\RR-\rr_i)]\mathbf
I\ ,\nonumber\\
&&\eta_{\rr\phi} = \sum_i[\eta(\rr_i) + \eta(\RR -
\rr_i)]d_i\mathbf t_i\ ,\nonumber\\
&&\eta_{\rr\RR} = -\sum_i\eta(\RR-\rr_i)\mathbf I\ ,
\label{170}
\end{eqnarray}
where $\mathbf I$ is a unit $2\times 2$ tensor.

In view of the symmetry (\ref{30}) of the dissipation coefficients,
the second of these equations (\ref{170}) uniquely fixes
$\eta_{\phi\rr} = \eta^T_{\rr\phi}$. To determine the remaining
coefficients $\eta_{\phi\phi}$ and $\eta_{\phi\RR}$, we
should consider the torque produced by the dissipative force as the
nanoparticle rotates between both planes. The magnitude of this torque
is $K_{diss} = -\sum_i d_i\left[\eta(\rr_i)\dot\rr_i +
\eta(\RR-\rr_i)(\dot\rr_i - \dot\RR)\right] \cdot \mathbf t_i$. Upon
substitution of the expression (\ref{130}) we find
\begin{eqnarray}
&&\eta_{\phi\RR} = -\sum_i\eta(\RR-\rr_i)d_i\mathbf t_i^T\
,\nonumber\\
&&\eta_{\phi\phi} = \sum_i\left[\eta(\rr_i) +
  \eta(\RR-\rr_i)\right]d_i^2\ .
\label{180}
\end{eqnarray}

For the upper plane, we have from the symmetry (\ref{30}) of the
dissipation coefficients $\eta_{\RR\rr} = \eta_{\rr\RR}$ and
$\eta_{\RR\phi} = \eta^T_{\phi\RR}$. The remaining tensor
$\eta_{\RR\RR}$ describes the effect of energy dissipation of the
upper plane into the internal degrees of freedom of the stationary
nanoparticle and has the form:
\begin{equation}
\eta_{\RR\RR} = \sum_i\eta(\RR-\rr_i)\mathbf I\ .
\label{190}
\end{equation}

\subsection{Overdamped zero-temperature limit}
It is difficult to estimate the dissipation coefficients from
first principles, because the basic building block of our model -- a
pseudoatom -- is a complex object consisting of many real atoms. It is not
unreasonable to assume though that the dissipation coefficient of such
a pseudoatom (and of the nanoparticle itself) can be many orders of
magnitude higher than that of a true atom on a surface. Therefore, in
our numerical calculations, we assume that the dissipation effects are
much stronger than the inertia effects, allowing us to consider the
overdamped limit by formally setting the nanoparticle's mass and
moment of inertia to zero: $m = 0$, $I = 0$. Likewise, we assume that
the spring attached to the upper plane is overdamped, allowing us to
set $M = 0$. Finally, since the potential energies from Eq.~(\ref{10})
represent an effect of many atoms, noise effects can be assumed
extremely small in comparison to the interaction forces and the normal
load. Therefore, we neglect thermal noise by setting $T$ to zero in
the equations of motion (\ref{40}), yielding
\begin{eqnarray}
&&\eta_{\rr\rr}\dot\rr + \eta_{\rr\phi}\dot\phi +
\eta_{\rr\RR}\dot\RR = \nonumber\\
&&\ \ \ \ \ -\nabla[ U(\rr,\phi) - U(\RR-\rr, \phi +
\pi)]\ ,\nonumber\\ 
&&\eta_{\phi\rr}\dot\rr +
\eta_{\phi\phi}\dot\phi + \eta_{\phi\RR}\dot\RR =\nonumber\\
&&\ \ \ \ \
-\frac{\partial[U(\rr,\phi)+U(\RR-\rr,\phi+\pi)]}{\partial\phi}\ ,
\nonumber\\ 
&&\eta_{\RR\rr}\dot\rr + \eta_{\RR\phi}\dot\phi +
\eta_{\RR\RR}\dot\RR = -\nabla U(\RR-\rr, \phi + \pi)\nonumber\\
&&\ \ \ \ \ \  - \eta_S(\dot\RR
- V\mathbf e_x) - \kappa(X - Vt)\mathbf e_x - f_N\mathbf e_y\ .
\label{200}
\end{eqnarray}

These equations can be simplified even further if we consider a
nanoparticle, which is symmetric with respect to rotations by $\pi$,
as in Fig.~\ref{fig1}. Due to this symmetry, and due to the
equivalence of the upper and lower planes, we can state that there is
a solution of Eqs.~(\ref{200}), for which
\begin{equation}
\rr = \RR/2
\label{210}
\end{equation}
up to an addition of an integer multiple of the lattice constant $a$
in the $x$-direction.
This is verified by inspection of the first of the equations of motion
(\ref{200}), where substitution of the relation (\ref{210}) renders
the force in the right-hand side vanish. Considering the left-hand
side, let us have a closer look at the damping coefficients from
Eq.~(\ref{170}). Because our nanoparticle is symmetric with respect to
rotations by $\pi$, for each pseudoatom at $\rr_i = \RR/2 + d_i\mathbf
e_i$ there is a symmetric partner pseudoatom at $\rr_k = \RR/2 -
d_i\mathbf e_i$. Then, comparison of the first and the third equations
(\ref{170}) yields $\eta_{\rr\rr} = -2\eta_{\rr\RR}$. Furthermore,
since the tangential vectors (\ref{140}) of the $i$th and the $k$th pseudoatoms
are opposite to each other, $\mathbf t_k = -\mathbf t_i$,
the sum in the second equation (\ref{170}) vanishes,
$\eta_{\rr\phi} = 0$, automatically implying that
$\eta_{\phi\rr} = 0$. Then, the first equation (\ref{200}) reduces
to $\eta_{\rr\RR}(\dot\RR - 2\dot\rr) = 0$, implying
Eq.~(\ref{210}). By numerically integrating the full set of equations
(\ref{200}), we have verified that the relation (\ref{210}) is stable:
for all initial conditions tried, the system eventually entered the regime
with $\rr = \RR/2$. Therefore, computational effort can be reduced by
roughly a factor of two by replacing five equations (\ref{200}) with
three equations of motion for $\phi, X$, and $Y$:
\begin{eqnarray}
&&\eta_{\phi\phi}\dot\phi + \eta_{\phi\RR}\dot\RR =
-2\frac{\partial U(\RR/2, \phi)}{\partial\phi}\ ,\nonumber\\
&&[\eta_S + \eta_{\RR\RR} + \eta_{\RR\rr}/2]\dot\RR +
\eta_{\RR\phi}\dot\phi = -\nabla U(\RR/2,
  \phi)\nonumber\\
&&\ \ \ \ + [\eta_S V\mathbf - \kappa (X-Vt)]\mathbf e_x - f_N\mathbf e_y\ .
\label{220}
\end{eqnarray}

\section{Results and Discussion}
\label{secIII}
\subsection{Parameters and units}
In all our numerical results below, we have chosen the lattice
constant $a$ as the unit of length, the adhesion energy $\varepsilon$
as the unit of energy, and the ratio $\eta_0 a^2/\varepsilon$ as the
unit of time. This choice fixes the unit of force to $\varepsilon/a$,
the unit of velocity to $\varepsilon/(a \eta_0)$, and the unit of
spring constant to $\varepsilon/a^2$. The value of the dissipation
coefficient at the minimum of the potential (\ref{100}) in these units
is $\eta_0 = 1$, and, obviously, $a = 1$, $\varepsilon = 1$ in these
units.

Our nanoparticle is constructed from an arrangement of pseudoatoms in
a square lattice with a lattice constant $b$, which, in general, is
not equal to the lattice constant $a$ of the two planes. From this
lattice we select those pseudoatoms whose distance from the center of
mass is smaller than the preset radius. In numerical simulations of
Eq.~(\ref{220}), we focused on an approximately round crystalline
nanoparticle with a radius of $5b$. The ``commensurate'' facets of the
nanoparticle correspond to the rotational angle given by an integer
multiple of $\pi/2$, $\phi_{comm} = n\pi/2$, whereas the
``incommensurate'' facets correspond to its half-integer multiple,
$\phi_{incomm} = (n + 1/2)\pi/2$.

In our simulations, we have tried different values of $b$, and
obtained qualitatively the same behaviour as for the $b = a = 1$ case
reported below.  As for other parameter values, we have taken the
equilibrium distance of the Lennard-Jones potential (\ref{100}) to be
equal the lattice constant, $\sigma = 1$. The corrugation depth of the
potential (\ref{110}) was taken to be $\Delta U = 3/4$, and
its decay length, as well as the decay length of the damping
coefficient (\ref{130}) were set to $\lambda = \xi = 1/5$. Finally,
the spring constant was set to $\kappa = 1$, and the
spring damping coefficient was $\eta_S = 10$. Other values of
these parameters of a comparable order of magnitude produced
qualitatively similar results.

\begin{figure}[h] 
\includegraphics[scale=0.8]{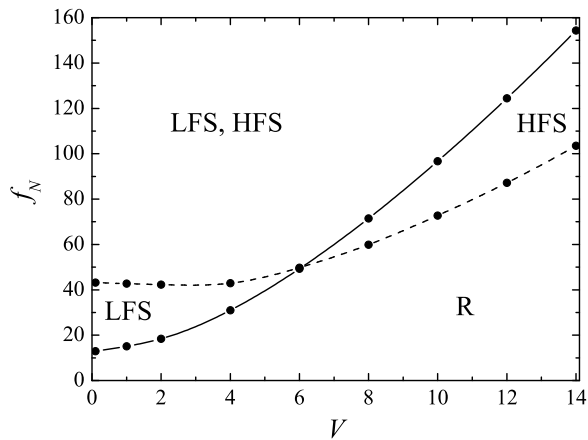}
\caption{State diagram of the system from Fig.~\ref{fig1} showing
  the stability regions of the low friction sliding (LFS), high
  friction sliding (HFS), and rolling (R) states.}
\label{fig2}
\end{figure}

\subsection{State diagram}
We have found that depending on normal load $f_N$ and pulling velocity
$V$, the nanoparticle can be rotating, or it can be stabilized in a
sliding state of either low or high friction. The results of our
numerical simulations are summarized in the state diagram of the
system, Fig.~\ref{fig2}, showing which friction regimes are stable for
given values of $f_N$ and $V$. When the normal load is sufficiently
low, the nanoparticle can only exist in the rolling (R) state of
motion, where the magnitude of the orientation angle steadily grows in
time, as in Fig.~\ref{fig4}(b). On the other hand, high normal load
stabilizes the sliding state of motion, which can be either the
low-friction sliding (LFS) or the high-friction sliding (HFS)
state. In both these states, the orientation angle $\phi$ of the
nanoparticle performs small rocking motion around the value, which is
either $\pi/4$ (LFS state) or $0$ (HFS state), up to an integer
multiple of $\pi/2$. The LFS state corresponds to the incommensurate
contact between the nanoparticle and the planes, while the
high-friction sliding (HFS) state corresponds to the commensurate
contact.  The LFS state is stable above the solid line in
Fig.~\ref{fig2}, the HFS state above the dashed line. Above both
separation lines, both LFS and HFS states are stable, and the actual
state of motion of the nanoparticle depends on its initial
preparation. Interestingly, to the left of the intersection point at
ca. $V = 6$, there is a region at lower normal loads where the LFS
state is the only stable state of motion; similarly, to the right of
this point, there is a region at higher normal loads where the
particle can exist only in the HFS state.

We now try to understand the state diagram from Fig.~\ref{fig2}. In
order to destroy the contact, work must be performed against the
adhesion forces and the normal load. This wok is done by the viscous
drag, which scales linearly with the velocity $V$, and by the force
generated by the moving corrugated potential. At low pulling
velocities, it is the latter force that is responsible for breaking
the contact. If the nanoparticle is in the LFS-state, the effective
corrugation of the potential is lower than that in the HFS-state, because
the contact is incommensurate and involves fewer atoms. Consequently,
a smaller normal load is required to stabilize the LFS state,
explaining its stability island in the low-velocity
region.

At faster pulling, on the other hand, the effect of the potential
corrugation becomes less important. This is so because the upper plane
moves relatively fast with respect to the particle, so that the
particle cannot follow the fast temporal variations of its potential
and feels, instead of the true corrugation depth $u_1$, a smaller
time-averaged corrugation. Therefore, it is the viscous drag that is
responsible for turning the nanoparticle at fast pulling. In order to
break the contact, one has to overcome the adhesion between the
nanoparticle and both planes. Since adhesion forces are larger in the
HFS-state, smaller normal load is required to stabilize this state at
fast pulling.

\begin{figure}[h] 
\includegraphics[scale=1]{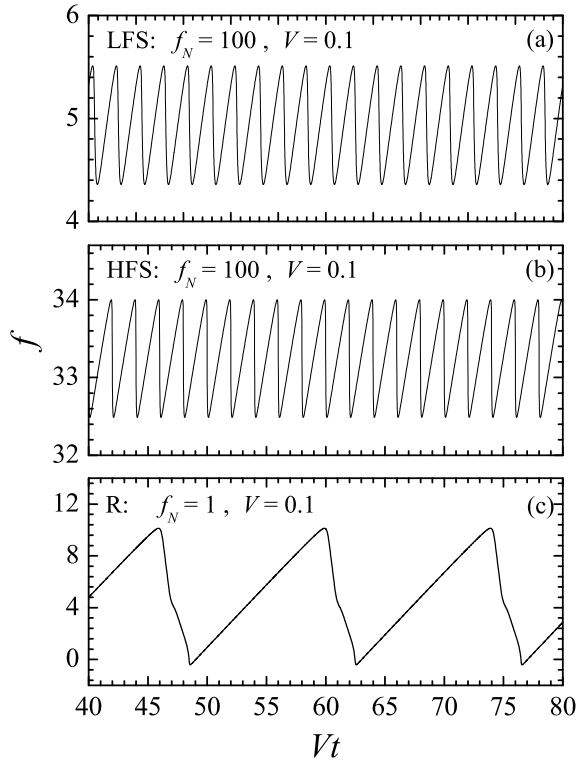}
\caption{Temporal evolution of the friction force (\ref{300}) in (a) the
  LFS-state, (b) the HFS-state, and (c) the R-state. All curves are
  obtained for the same pulling velocity $V = 0.1$,
  but different normal loads: $f_N = 100$ for the curves (a)
  and (b), and $f_N = 1$ for the curve (c). The difference
  between the LFS- and HFS-curves (a) and (b) is in the orientation
  angle of the nanoparticle: namely, for the curve (a), the angle
  $\phi$ is close to $\pi/4$, and for the curve (b), it is close to zero.}
\label{fig3}
\end{figure}

\subsection{Friction force}
Fig.~\ref{fig3} exemplifies the typical evolution of the friction
force (\ref{300}) in the three states. The curves (a) and (b) resemble
the typical evolution of the friction force in an atomic friction
experiment in the stick-slip regime \cite{Mate87, Evst06}. During the stick
phases, the nanoparticle and the upper plane are almost stationary,
while the elastic deformation of the spring attached to the top plane
constantly increases due to pulling. When elastic energy becomes
sufficient to initiate the slip, the nanoparticle gets displaced in
the $x$-direction by one lattice constant, and the upper plane by two
lattice constants, see Eq.~(\ref{210}), resulting in a sudden
relaxation of the spring. Consequently, the periodicity of the
stick-slip curves in the LFS regime (a) and the HFS regime (b) equals
two lattice constants. We note that the modulation amplitude of the
stick-slip curves (a) and (b) is about the same, while the mean
friction forces developed in both regimes are very different, in spite
of the fact the pulling velocity and the normal loads are identical
for the curves (a) and (b).

The curve (c) depicting the evolution of the friction force in the
rolling state is interesting in three respects: its periodicity is
notably larger than in the LFS- and HFS-cases (a) and (b); its
modulation amplitude is also surprisingly high; the slope of
the curve in a stick phase is much smaller than in the cases (a)
and (b). These peculiarities can be explained as follows. 

The periodicity of the ``stick-roll'' curve (c) is
equal to the distance travelled by the upper plane during the ``slip''
\cite{Evst06}. This distance is quite large, because the ``slip''
event, seen as the sudden drop of the friction force, is associated
not with the transition of the nanoparticle by one lattice constant,
but with its rotation by an angle $\pi/2$ from one commensurate
contact state to the next. During this rotation, the particle's center
of mass travels a distance of about $\pi R/2$. The actual distance is
slightly smaller than this value, because the nanoparticle is not
perfectly round. In view of Eq.~(\ref{210}), the periodicity of the
curve (c) twice that value, that is, slightly smaller $\pi R$. For $R
= 5$, this gives the periodicity that should be a bit smaller than 16
lattice constants. The periodicity of the curve (c) is indeed 14
lattice constants.

The second interesting feature is a roughly ten-fold larger amplitude of
force variations in the rolling state compared to the LFS- and
HFS-curves in Fig.~\ref{fig3}, in spite of the fact that the normal
load is two orders of magnitude smaller. As it turns out, the reason
is precisely the much smaller normal load. In each stick phase, the
elastic energy of the spring constantly builds up and is suddenly
released to induce a rotation of the nanoparticle in the rolling state
from Fig.~\ref{fig3} (c). At the same time, the nanoparticle in the
stick phase can slightly turn and lift the top plane up against the
small normal load. Since part of the torque applied to the particle by
the spring is used to lift the upper plane, a much larger elastic
force is necessary to induce the particle's rotation, implying a large
modulation amplitude of the curve in Fig.~\ref{fig3}(c). For the
curves from Fig.~\ref{fig3} (a) and (b), on the other hand, the normal
load is too high to allow for any significant lifting of the upper
plane, so that practically all of the force accumulated in the stick
phase is used to initiate the slip.

Finally, the observed rate of force increase in the stick phase of the
rotational regime (c) is much smaller than in the sliding regimes (a)
and (b) for a similar reason. The rate of force increase in the
stick phase is determined by an effective spring constant,
$\kappa_{eff}$, which is given by a combination of the elasticity of
the spring $\kappa$ attached to the upper plane and the spring
constant of the nanoparticle's contacts with both planes,
$\kappa_{cont}$. Since this ``contact'' spring is attached to the
spring of the upper plane in series, the combination rule is
$1/\kappa_{eff} = 1/\kappa + 1/\kappa_{cont}$, meaning that
$\kappa_{eff} < \kappa_{cont}$. In view of the large difference in the
normal load, the contact in the sliding regimes (a) and (b) is much
more rigid than in the rolling regime (c). This implies a much smaller
effective spring constant in the case (c) than in the cases (a) and
(b), and a smaller rate of force increase.

\begin{figure}[h] 
\includegraphics[scale=1]{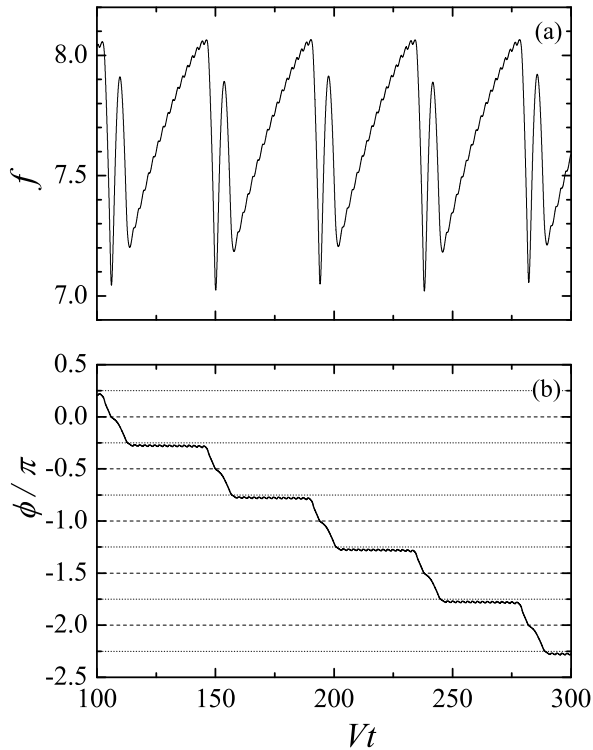}
\caption{Temporal evolution of (a) the friction force $f$ from Eq.~(\ref{300})
  and (b) the rotational angle $\phi$ in the rolling state
  corresponding to the pulling velocity $V = 4$ and normal load $f_N =
  30$. }
\label{fig4}
\end{figure}
\begin{figure}[h] 
\includegraphics[scale=1]{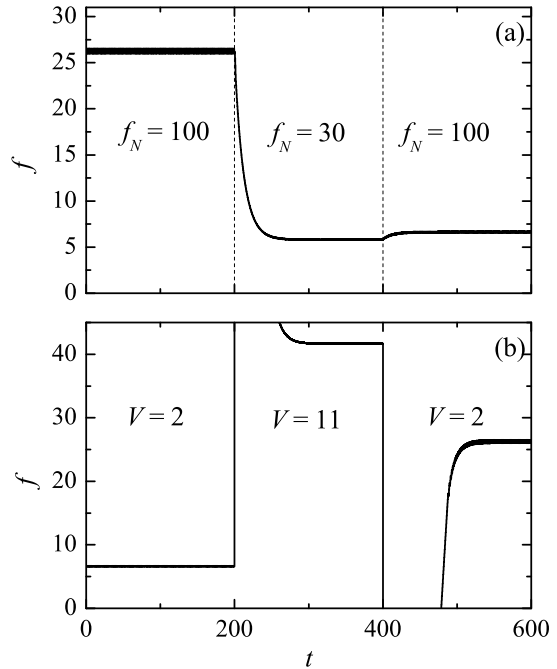}
\caption{Evolution of the friction force $f$ from Eq.~(\ref{300})
  during switching of the friction regime: (a) switching from HFS to
  LFS is performed by means of applying a negative pulse of the normal
  load at a constant velocity $V = 2$; (b)
  switching from LFS to HFS is achieved by applying a positive
  velocity pulse at a constant normal load $f_N = 100$.}
\label{fig5}
\end{figure}

Fig.~\ref{fig4} shows that in the rolling state, the shape of the
force curve [Fig.~\ref{fig4}(a)] can be quite different from the
saw-tooth-like ones shown in Fig.~\ref{fig3}. Here, the rotation angle
[Fig.~\ref{fig4}(b)] in the stick phase has the value $\phi = n\pi/2 +
\pi/4$, meaning that the nanoparticle contacts the planes along its
``incommensurate'' facets. In the end of such a stick phase, the
nanoparticle first quickly rotates by an angle of $\pi/4$ and enters
another short-lived stick phase, where the contact is formed along the
``commensurate'' facets of the nanoparticle. Then, another rotation
into an incommensurate state occurs, and the process repeats
itself. As a result, the force evolution curve acquires an additional
structure, with different stick phases corresponding to different
particle-surface contacts. We note that the force curve observed in
the carbon nanotube rolling experiments \cite{Falvo99} also possesses
a rather complex structure, presumably because of the variation of the contact
properties of the nanotube ``facets''.

\subsection{Control of friction}
The coexistence of the LFS and HFS states in Fig.~\ref{fig2} opens the
possibility of switching between them by changing the control
parameters, $f_N$ and $V$. This process is illustrated in
Fig.~\ref{fig5}. Starting with the HFS-state within the coexistence
region at $f_N = 100$ and $V = 2$, we
reduce the normal load to $f_N = 30$ keeping the velocity
constant, Fig.~\ref{fig5}(a). This brings the system into that region
of the state diagram from Fig.~\ref{fig2} where the LFS-state is the only
stable one and induces a rotation of the nanoparticle by
$\pi/4$. Restoring the normal load to its initial value brings the
nanoparticle back into the coexistence region, but now its contact
with both planes goes along the incommensurate sides.

In order to switch the nanoparticle back into the HFS state, one can
increase the pulling velocity to a larger value, thus bringing the
system out of the coexistence region into the HFS part of the state
diagram, see Fig.~\ref{fig5}(b). This results in a sudden stretching
of the spring and, correspondingly, in a large spike of the elastic
force. After the nanoparticles has rotated into the HFS configuration,
the velocity is reduced to the initial value in the coexistence region
of the state diagram. Again, this velocity reduction results in a
sudden relaxation of the spring and in the large negative spike in the
friction force. After that spike, the friction force stabilizes at a
high value corresponding to the HFS-state of the system.

\section{Conclusions}
To summarize, we have considered the rolling and sliding motion
regimes of a cylindrical crystalline nanoparticle between two crystalline
planes and found that the rolling state of motion is stable for
sufficiently weak normal loads and fast pulling. An intriguing feature
of the state diagram from Fig.~\ref{fig2} is the existence of the
stability island of the LFS state, where the contact between the
nanoparticle and the planes is incommensurate.  This is in striking
contrast with the finding of Refs.~\onlinecite{Depondt98, Filippov08}, where
the HFS commensurable state was the only stable one. The reason for
this difference is that, in the works \onlinecite{Depondt98, Filippov08},
the flat nanoparticle was rotating around the axis perpendicular to
the planes, while in our work, the rotation axis is parallel to the
planes and perpendicular to the direction of motion. We have shown
that using nanoparticles as a lubricant, one can achieve either low or
high friction at the same values of normal load and pulling
velocity. This finding is at variance with the results from
Refs.~\onlinecite{Socoliuc04, Krylov05, Socoliuc07}, where adjustment of
these parameters was essential for friction control, so that different
friction regimes could be realized only for different values of $V$
and $f_N$.

Even though the results reported in this paper have been obtained for
the special case of equal lattice constants of the nanoparticle and
the planes, we have found qualitatively similar behaviour also in the
case of unequal lattice constants. The fact that some facets of the
nanoparticle are commensurate or incommensurate with the surfaces
facilitates the discussion of the results, but is not central for the
physical mechanism underlying the friction states of the
nanoparticle. What is important is that different facets of the
nanoparticle are characterized by different interaction energies with
the surfaces. If this condition is fulfilled, then several sliding
states and the rolling state can be realized, even when all facets are
incommensurate with the surfaces, or if they are commensurate, but
differ in the number of contact atoms.

In our analysis, several important effects have been neglected, such
the nanoparticle's possible asymmetry, inertia effects, and the effect
of thermal noise. All these effects, when properly taken into account,
may lead to qualitatively new friction regimes. In particular,
nanoparticles of other shapes can exhibit different state diagrams
and, correspondingly, can allow for different friction switching
mechanisms. For instance, if the nanoparticle is asymmetric, it can
stick to one plane and slide against the other; or it can slide with
respect to both planes, but with the velocity very different from half
the velocity of the upper plane. Next, the effects of thermal noise
are typically negligible compared to load and interaction
forces. However, in the systems where noise effects play a significant
role, the nanoparticle is expected to spontaneously perform thermally
induced transitions between the LFS and the HFS states within the
coexistence region, so that higher normal loads would be required to
stabilize them. In this case, one can expect the appearance of new
regions in the state diagram, where friction is controlled by thermal
noise. Finally, inertia introduces new characteristic time scales into
the problem -- the inverse resonance frequencies of the nanoparticle's
vertical, horizontal, and angular oscillations within the potential of
the two surfaces. If the time of pulling by one lattice constant,
$a/V$, becomes comparable to any of these time scales, new friction
regimes associated with the nanoparticle's resonant motion can be
expected. Exploring new friction regimes related to these and possibly
other factors can be an exciting subject for future research, both
theoretical and experimental.\\

\section{Acknowledgements}
We thank the Deutsche Forschungsgemeinschaft (Collaborative Research
Center SFB 613) and the ESF programs NATRIBO and FANAS (collaborative
research project Nanoparma -- 07-FANAS-FP-009) for financial support.

\end{document}